\documentclass[fleqn,twoside]{article}
\usepackage{espcrc2}
\usepackage{amsmath}

\usepackage{graphicx}
\usepackage[figuresright]{rotating}

\newcommand{\be}{\begin{equation}}
\newcommand{\ee}{\end{equation}}
\newcommand{\lwig}{\mbox{\,\raisebox{.3ex}
    {$<$}$\!\!\!\!\!$\raisebox{-.9ex}{$\sim$}\,}}

\title{
\vspace*{-1.2cm}{\normalsize\rightline{ITP-BUDAPEST 602}
\rightline{DESY 03-164}\rightline{hep-lat/0310051}}
Lattice QCD at finite $T$ and $\mu$}

\author{S.D. Katz\address[DESY]{Deutsches Elekronen-Synchrotron 
        DESY, Hamburg, Germany}%
        \thanks{On leave from Institute for Theoretical Physics, 
        E\"otv\"os University, Budapest, Hungary}}

\begin{document}

\begin{abstract}
Recent results of lattice QCD at finite temperature and density are reviewed.
At vanishing density the transition temperature, the equation of state
and hadron properties are discussed both for the pure gauge theory and for 
dynamical staggered, Wilson and overlap fermions. The second part deals with
finite density. There are recent 
results for full QCD at finite temperature and 
moderate density, while at larger densities QCD-like models are studied.
\vspace{1pc}
\end{abstract}

\maketitle

\section{INTRODUCTION}
Since the first phase transition was seen in lattice gauge theories 
in 1981 
\cite{McLerran:1980pk,Kuti:1980gh}, lattice QCD at finite temperature
has grown into a large field becoming close to give real, 
continuum-extrapolated predictions for full QCD, which is essential for 
understanding phenomena of the early universe and at heavy-ion experiments.
As perturbation theory is only reliable at very high temperatures, lattice 
QCD seems to be the only tool to answer such questions as the transition 
temperature between hadronic matter and the quark-gluon plasma, the 
equation of state of strongly interacting matter or hadron properties
at finite temperature and/or density.

\subsection{Improved actions}
The formulation of QCD on the lattice is not unique. One has different
choices for the lattice action which all give the same continuum limit.
The simplest choices are the Wilson action in the gauge sector and the
Wilson or staggered action for fermions. Using these actions, however 
may lead to large cutoff effects unless we use fine enough lattices.
It is often argued, that a good compromise is to use improved actions,
which are computationally (much) more expensive, but may 
significantly reduce cutoff effects. 
There are many possibilities for improved actions both in the gauge and
fermionic sectors.

One has to keep in mind, however, that for 
reliable results, no matter what action
is used, simulations at least at two different lattice spacings 
(thus different
temporal extensions ($N_t$) ), and a corresponding 
continuum extrapolation is required.
As going to larger $N_t$ is often very costly using improved actions, this 
fact may easily lead to the conclusion that using the simplest standard
actions and a few different $N_t$-s may give more precise results than using
just one single $N_t$ with an improved action.

\begin{figure*}
\centerline{
\includegraphics[width=7.0cm,height=6cm]{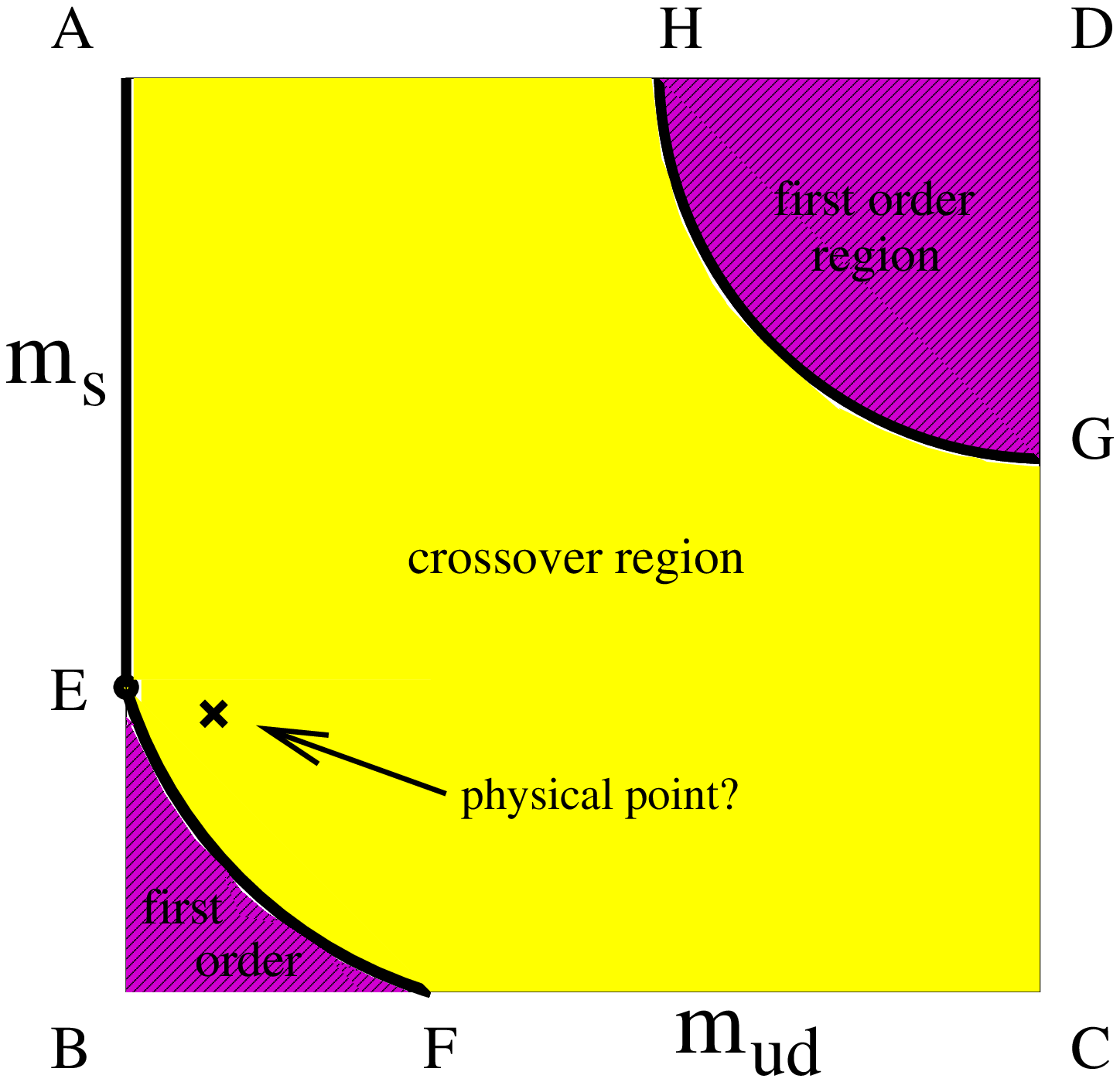}\hspace*{1.5cm}
\includegraphics[width=7.0cm]{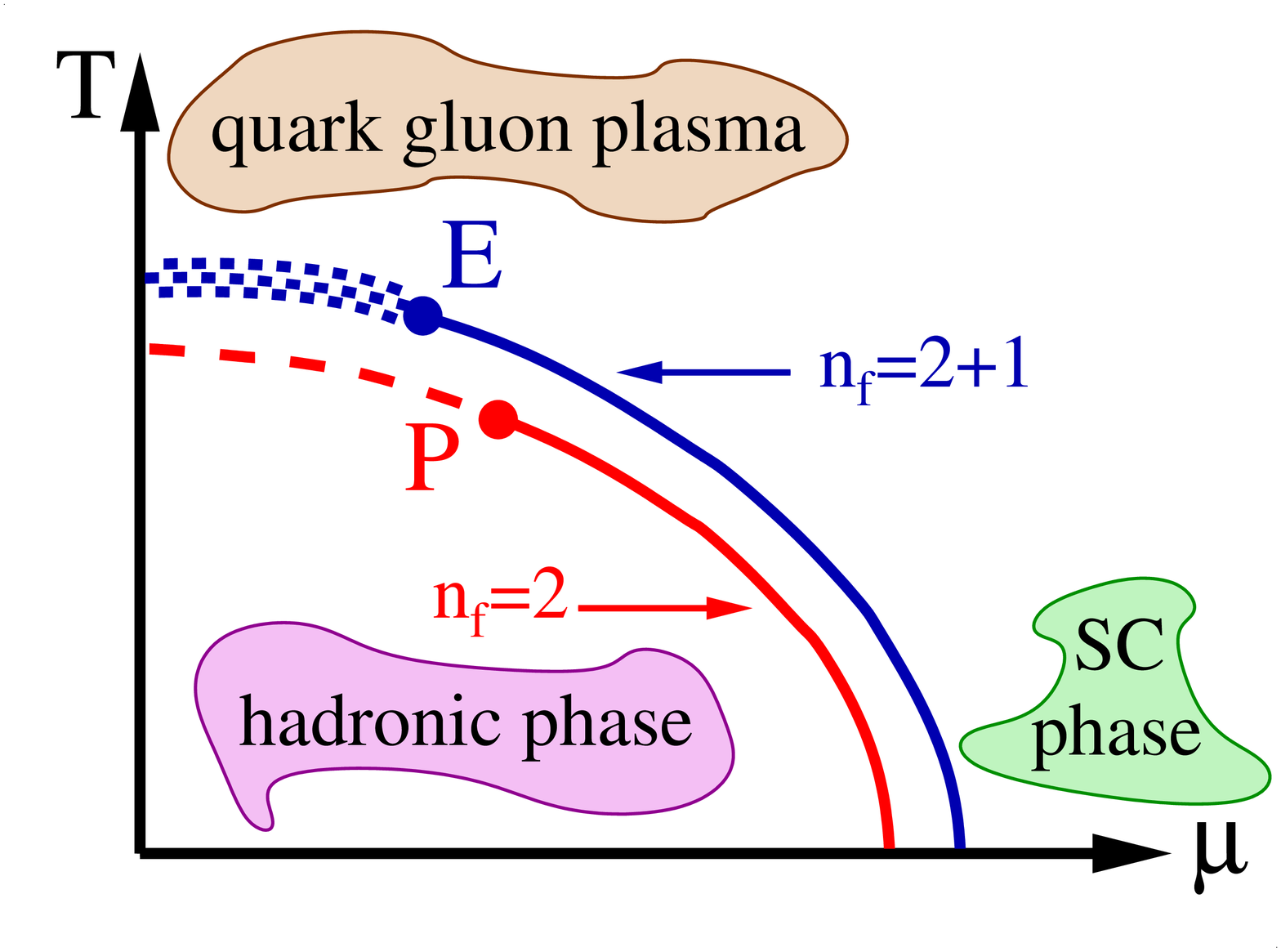}
}\vspace*{-0.6cm}
\caption{Left: the QCD phase diagram in the $m_{ud}, m_s$ plane 
The thick solid lines represent second order transitions separating
cross-over regions from first order ones.
Right: the QCD phase diagram in the $\mu, T$ plane. In the case of
two massless flavors, a second order region (dashed line) is connected
to a first order region via the tricritical point ($P$). In the 
realistic case of two light and one heavier flavor, a crossover (dotted
region) is connected to the first order region via the critical endpoint ($E$).}
\label{pd}
\end{figure*}
\subsection{The QCD phase transition}
At high temperature ($T$) and/or large chemical potential ($\mu$)
the hadronic matter undergoes a transition
into a quark-gluon dominated phase. At zero $\mu$ this transition happens
at $T\approx 270$~MeV in the quenched case while at $T\approx 150-170$~MeV
in full QCD. In the pure gauge case the action has an exact $Z(3)$ symmetry
which is spontaneously broken in the low temperature phase. The Polyakov 
loop ($L$) is not invariant under this symmetry, thus it is an order parameter of the
phase transition. In massless dynamical QCD we have chiral symmetry, which 
is expected to be broken in the low temperature phase. Correspondingly, the
chiral condensate behaves as an order parameter in this case.
In the realistic case with nonzero quark masses none of these symmetry 
breakings characterize the transition, nevertheless usually both the 
Polyakov loop and the chiral condensate shows a rapid change 
(or discontinuity) around the transition.

The transition can be usually located by observing the change of these 
order parameter-like quantities. Usually their susceptibilities show a nice
signal which can be used as a definition of the transition point.
An alternative possibility is to locate the Lee-Yang zeroes of the partition
function~\cite{Yang:be}. This gives usually more precise values for $T_c$.

An interesting and natural question is, whether the transition is of 
first order, second order or maybe only a rapid cross-over. As lattice
simulations are always done on lattices with a finite volume ($V$)
where no singularities, thus
no real phase transitions may occur, the only way to determine the order
of the transition is to examine the finite size scaling behavior of certain
observables, like the susceptibility peaks, Binder cumulants or the imaginary
parts of Lee-Yang zeroes.

The phase diagram of QCD in the plane of the quark masses
is shown in Fig. \ref{pd} (left). Based on universality arguments, 
one expects a first order phase transition for three flavors of massless
quarks and a second order one for two massless flavors. In the quenched 
limit with infinite quark masses the transition is of first order as well.
At nonzero, but finite quark masses the picture is more complicated.
For intermediate masses we expect only a crossover. Whether the physical 
point lies within the first-order or cross-over region, is still an open 
question, however the cross-over is much more likely.

In the latter case an interesting picture emerges if we introduce a nonzero
chemical potential ($\mu$). At zero temperature and large enough $\mu$ one
expects a strong first order phase transition. If there is only a cross over
at zero $\mu$ and finite $T$ then there should be a critical endpoint
on the line connecting the two regions as seen in
Fig. \ref{pd} (right). This endpoint is an unambiguous
non-perturbative prediction of the QCD Lagrangian and has important 
experimental signatures~\cite{Stephanov:1998dy}.

\begin{figure*}
\centerline{
\includegraphics[width=7.2cm]{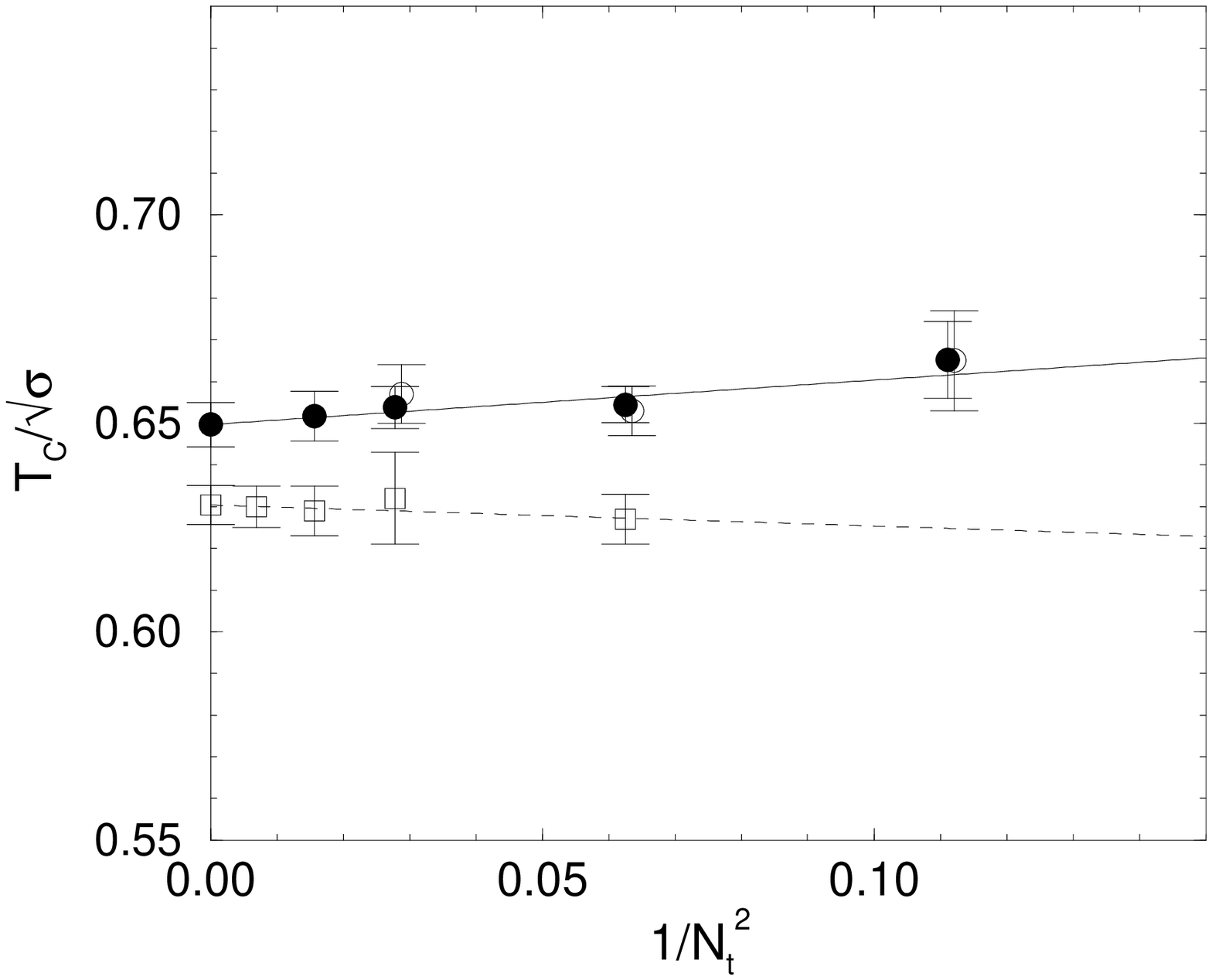} \hspace*{1.3cm}
\includegraphics[width=7.2cm]{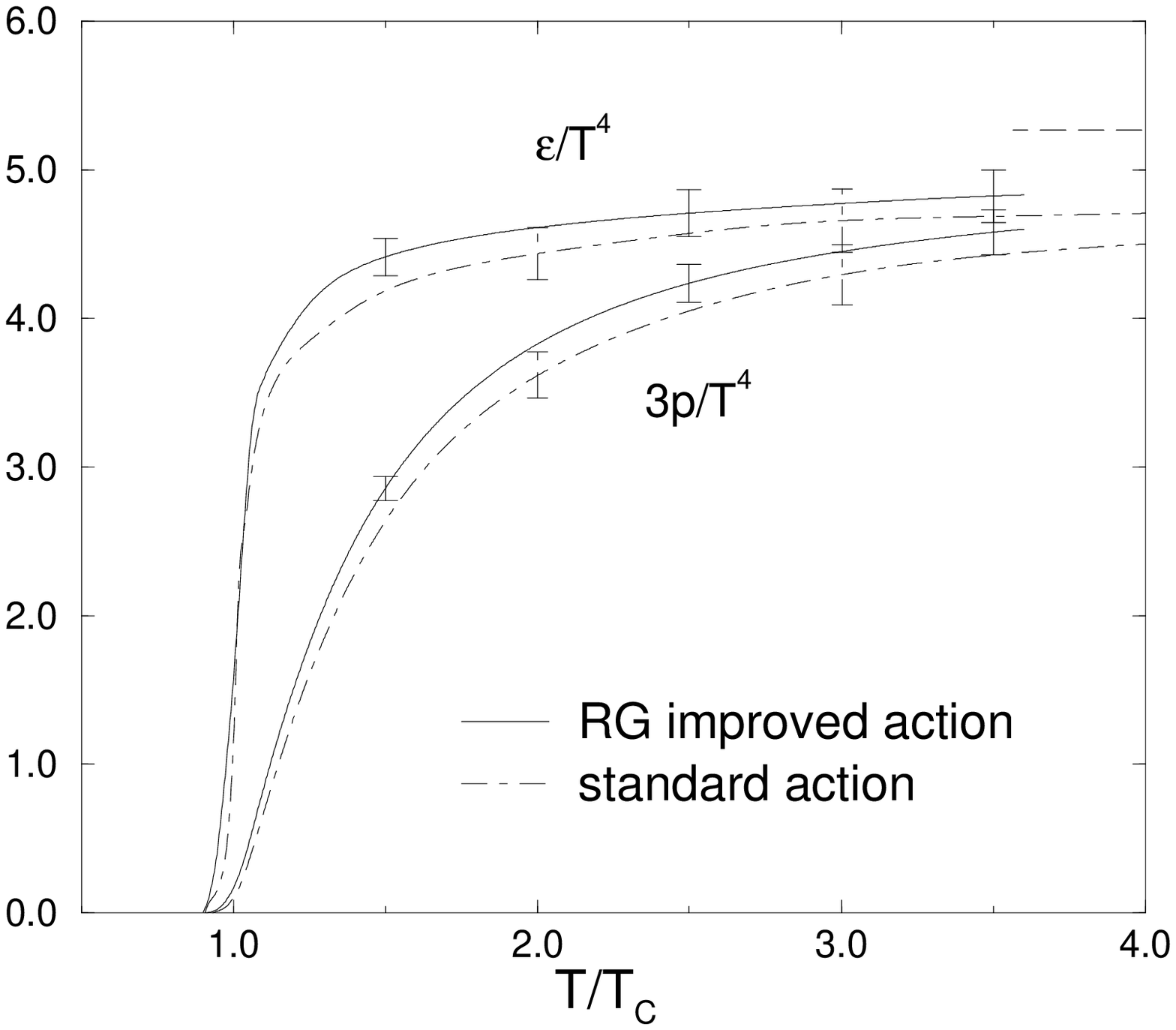}
} \vspace*{-0.6cm}
\caption{Left: the continuum extrapolation of $T_c/\sqrt{\sigma}$ in the
pure SU(3) theory using standard and RG-improved actions. Right:
the pressure and energy density as a function of the temperature for 
standard and RG-improved actions.}
\label{pg}
\end{figure*}

\subsection{The equation of state along the line of constant
physics}
Besides finding the transition temperature (which may be 
not even well defined in the case of a cross-over) 
and/or giving the phase diagram in the $\mu-T$ plane, one can 
also give the equation of state (EoS) of hadronic/gluonic matter well below or 
above the transition $T$,or $\mu$. All thermodynamical quantities 
can be obtained from the grand canonical partition function, e.g. for the
energy density and the pressure:
\be
\epsilon (T)=\frac{T^2}{V}\frac{\partial(\log Z)}{\partial {T}} \;\;\;
p(T) = T\frac{\partial (\log Z)}{\partial V}.
\ee
On lattices with isotropic couplings
we can not take derivatives with respect to
$T$ and $V$ independently, as they are connected via the lattice 
spacing ($a$). The combination $\epsilon-3p$ can be written using
derivatives with respect to $a$:
\be
\frac{\epsilon-3p}{T^4}=-\frac{L_t^3}{L_s^3}\left\{ a \frac{d (\log Z)}{d a}
-\mu \frac{d(\log Z)}{d \mu} \right\},
\ee
which can be rewritten in terms of the Plaquette ($P$) and 
chiral condensate ($\bar{u} u$ and $\bar{s} s$) expectation values:
\begin{eqnarray}
\frac{\epsilon-3p}{T^4}=
-L_t^4 & \left[ a\frac{\partial \beta}{\partial a} <{\rm P}>+
a\frac{\partial (a m_u)}{\partial a} <\bar{u}u> \right. \nonumber \\&+
\left.
a\frac{\partial (a m_s)}{\partial a} <\bar{s}s>\right].
\end{eqnarray}
Here all derivatives should be taken along the line of constant physics (LCP)
where the physical values of the quark masses (and not the $am_i$ lattice
masses) are kept constant. In order to determine both $\epsilon$ and $p$
one needs another combination, which can be the pressure itself. For
large homogeneous systems, $p\propto \log Z$, which can be computed using the 
integral method:
\begin{eqnarray}
\frac{p}{T^4}=
L_t^4\int d\beta \left[<{\rm P}>+
m_u\frac{\partial a}{\partial \beta}<\bar{u}u> \right.\nonumber \\ +\left.
m_s\frac{\partial a}{\partial \beta}<\bar{s}s>+
\mu\frac{\partial a}{\partial \beta}\frac{\partial(\log \det M)}{\partial (\mu a)}\right].
\end{eqnarray}
Here again the integration should follow the LCP and the subtraction
of $T=0$ observables are assumed.

In the rest of this review recent results of lattice QCD at finite $T$ 
and/or $\mu$ are summarized. In the next section 
quenched results are reviewed. Sect. \ref{finT} deals with results of
dynamical simulations using Wilson, staggered or overlap fermions.
In sect. \ref{finmu1} the recent progress of exploring the
$\mu\neq0$ regions of the QCD phase diagram is discussed. Sect. \ref{finmu2}
shortly introduces some QCD-like models used for studying large chemical
potentials and finally, sect. \ref{conc} concludes.
\begin{figure*}
\centerline{
\includegraphics[width=7.0cm]{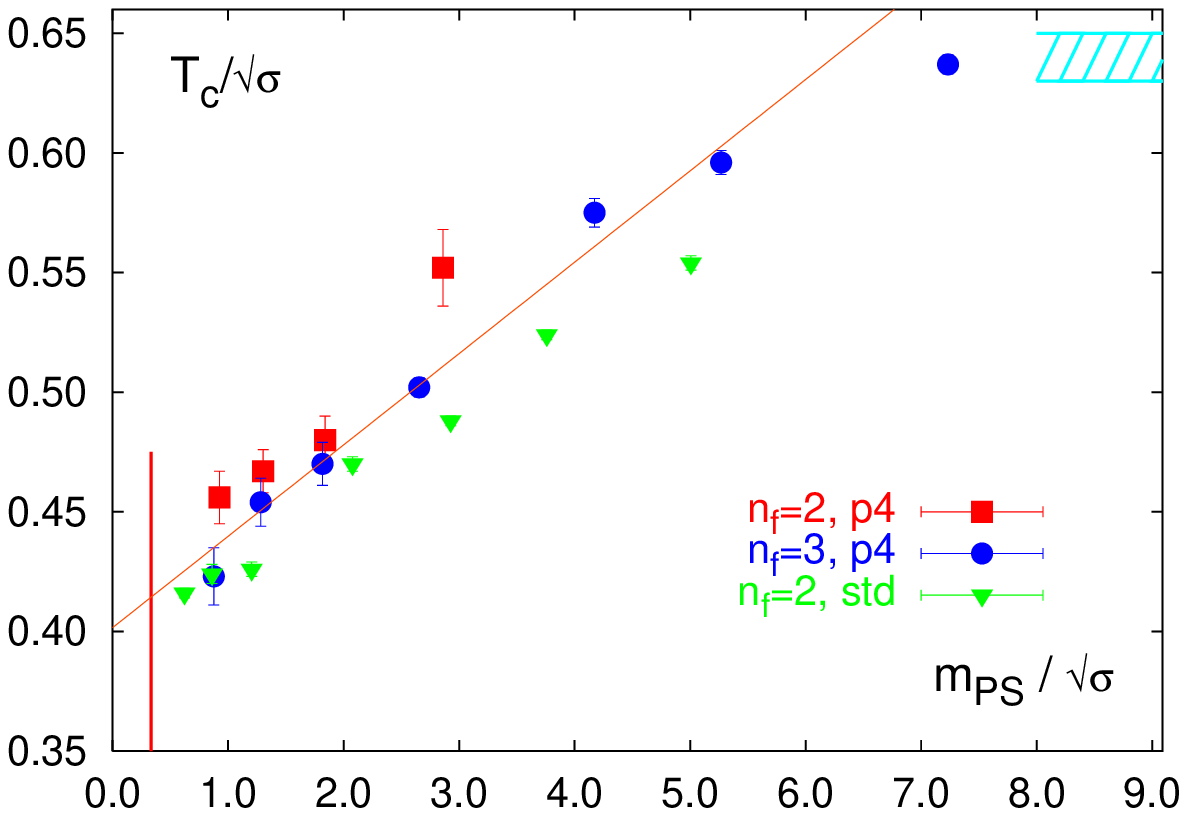} \hspace*{1.6cm}
\includegraphics[width=6.0cm]{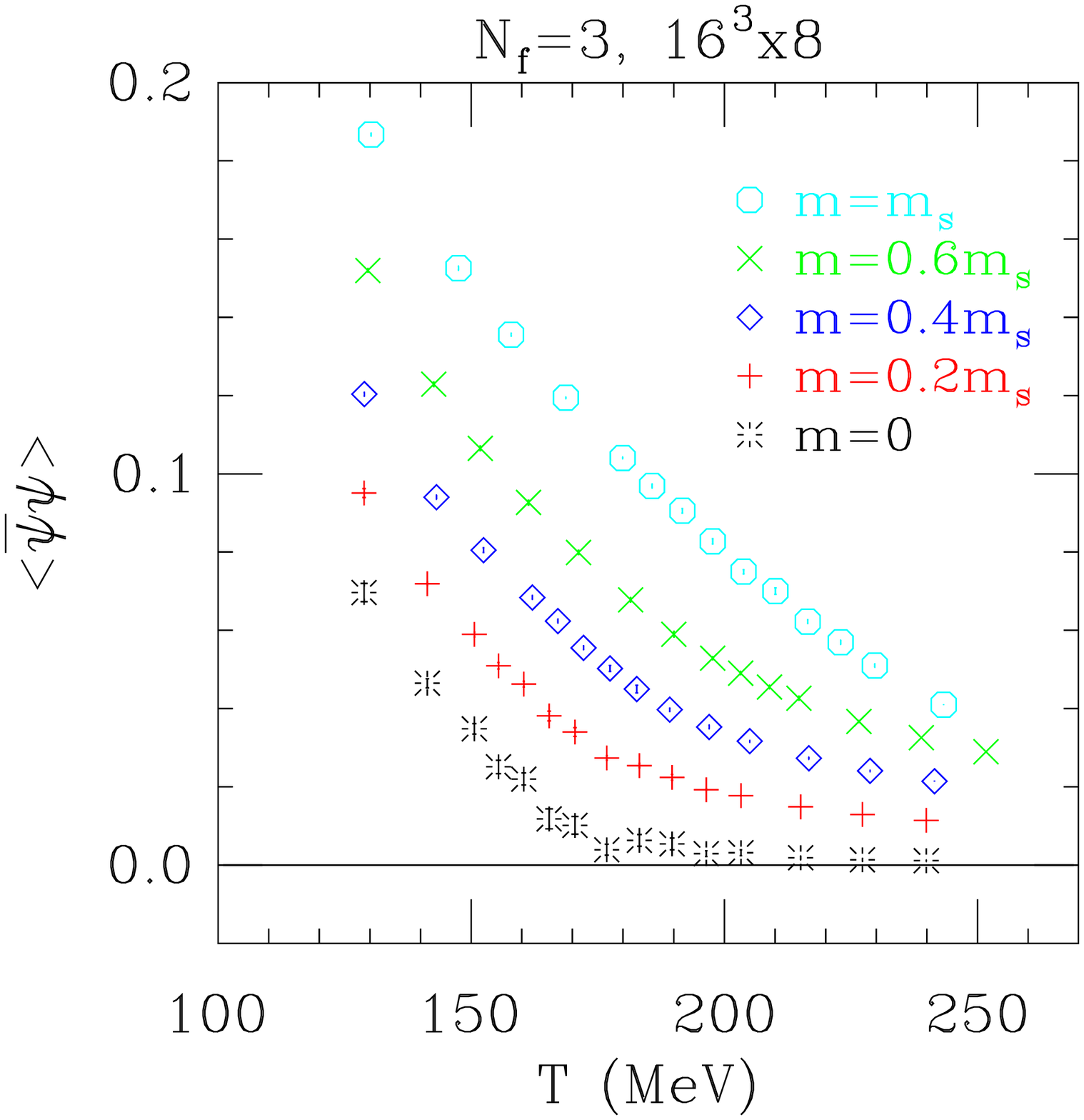}
}\vspace*{-0.6cm}
\caption{Left: the chiral extrapolation of $T_c/\sqrt{\sigma}$ for 
$N_f=2$ (squares), 3 (circles) for the $p4$ and $N_f=2$ (triangles) for
the standard staggered action.
Right: the chiral condensate as a function of the temperature
for several quark masses and extrapolated to $m=0$ with $N_f=3$ 
ASQTAD fermions.
}
\label{stag}
\end{figure*}

\section{PURE GAUGE RESULTS}\label{s_pg}
The simplest case to study is pure SU(3) gauge theory without dynamical
quarks. Using both standard~\cite{Boyd:1996bx} and improved 
actions~\cite{Beinlich:1997ia,Okamoto:1999hi}
on isotropic and anisotropic~\cite{Namekawa:2001ih} lattices
the critical temperature has been determined in the continuum limit (see 
Fig.~\ref{pg} left).
The results are usually given in units of the string tension and 
one gets
$T_c/\sqrt{\sigma}\approx 0.630-0.650$.
The results seem to be not completely consistent, however the difference
comes from possible systematic errors in measuring the string tension.
If one sets the scale using the $r_0$ parameter~\cite{Sommer:1993ce}
then a perfect agreement is found~\cite{Necco:2003vh}.
The equation of state is shown on Fig.~\ref{pg} (right) for both actions.
As the only parameter here is the gauge coupling which is used to set the
scale, the system is kept automatically on the LCP.
We can see that even at temperatures as high as $4T_c$ there is still 
15\% deviation from the Stefan-Boltzmann (SB) free gas limit.
A careful finite size scaling analysis shows that the transition is of
first order as expected~\cite{Fukugita:1989yb}.

\section{DYNAMICAL RESULTS AT $\mu=0$} \label{finT}

Including dynamical quarks increases the cost of simulations by at least
two orders of magnitude. Another difficulty is introduced by the quark 
masses. As direct simulations at the physical values of the u,d quark masses
are still practically impossible, chiral extrapolation is usually performed
from larger masses. Taking the continuum limit is very expensive 
and it is usually not yet done. Therefore we still have to wait for 
final results.

\subsection{Staggered results for $T_c$ and the EoS}

The staggered discretization of the fermionic action describes four flavors
of quarks with equal masses. Taking a fractional power of the fermion 
determinant one can also introduce less flavors. Using staggered fermions 
is significantly faster than Wilson fermions and another advantage is that
part of the chiral symmetry is preserved.
Fig. \ref{stag} (left) shows the results of the Bielefeld-group on $T_c$
using $p4$ action and $16^3\times 4$ lattices~\cite{Karsch:2000kv}. 
The critical points were defined by
susceptibility peaks. The critical temperature on these $N_t=4$ lattices
in the chiral limit is 
$T_c\approx 150(170)$~MeV for two (three) flavors. 
The MILC collaboration has somewhat different results using the ASQTAD 
action with 2+1 and 3 flavors~\cite{Bernard:2003dk}
(Fig.~\ref{stag} right).
On $16^3\times 8$ lattices they see a weak crossover at somewhat larger
$T_c \approx 180$~MeV. Note, that although they have a larger $N_t$, the 
volume they use is much smaller than the one used by the Bielefeld group.
Thus, there may still be significant finite-size effects.
Clearly, in order to have a reliable prediction for $T_c$ one has 
to take the thermodynamical limit and carry out
a continuum extrapolation.

Both of these analyses used degenerate $u,d$ quark masses. It has been 
investigated in \cite{Gavai:2002fi} that breaking the flavor SU(2) symmetry 
and using somewhat different $u,d$ masses does not change $T_c$ significantly.
\begin{figure}
\centerline{
\includegraphics[width=8.0cm]{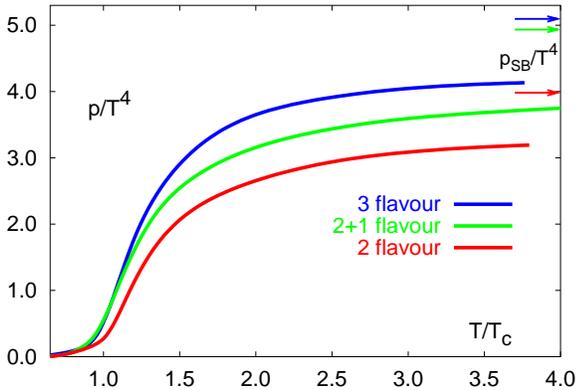}
}\vspace*{-0.6cm}
\caption{
The pressure as a function of $T/T_c$ for $N_f=3,2+1,2$
(from top to bottom) $p4$ fermions.}
\label{eos_stag}
\end{figure}

Using standard staggered action the equation of state has been determined
by the MILC collaboration~\cite{Bernard:1996cs} on $N_t=4$ and 6.
Fig.~\ref{eos_stag} shows the pressure obtained by the Bielefeld group with $p4$ 
action on $16^3\times 4$ 
lattices~\cite{Karsch:2000ps}.

Note that both analyses kept $a m_q$ constant which is not an LCP approach
as discussed in the introduction. In these cases the physical quark masses
increase as the temperature is increased. 
This leads to a 10-15\% suppression of the pressure at high temperatures
compared to the LCP approach~\cite{eos_long}.

\begin{figure*}
\centerline{
\includegraphics[width=7.2cm]{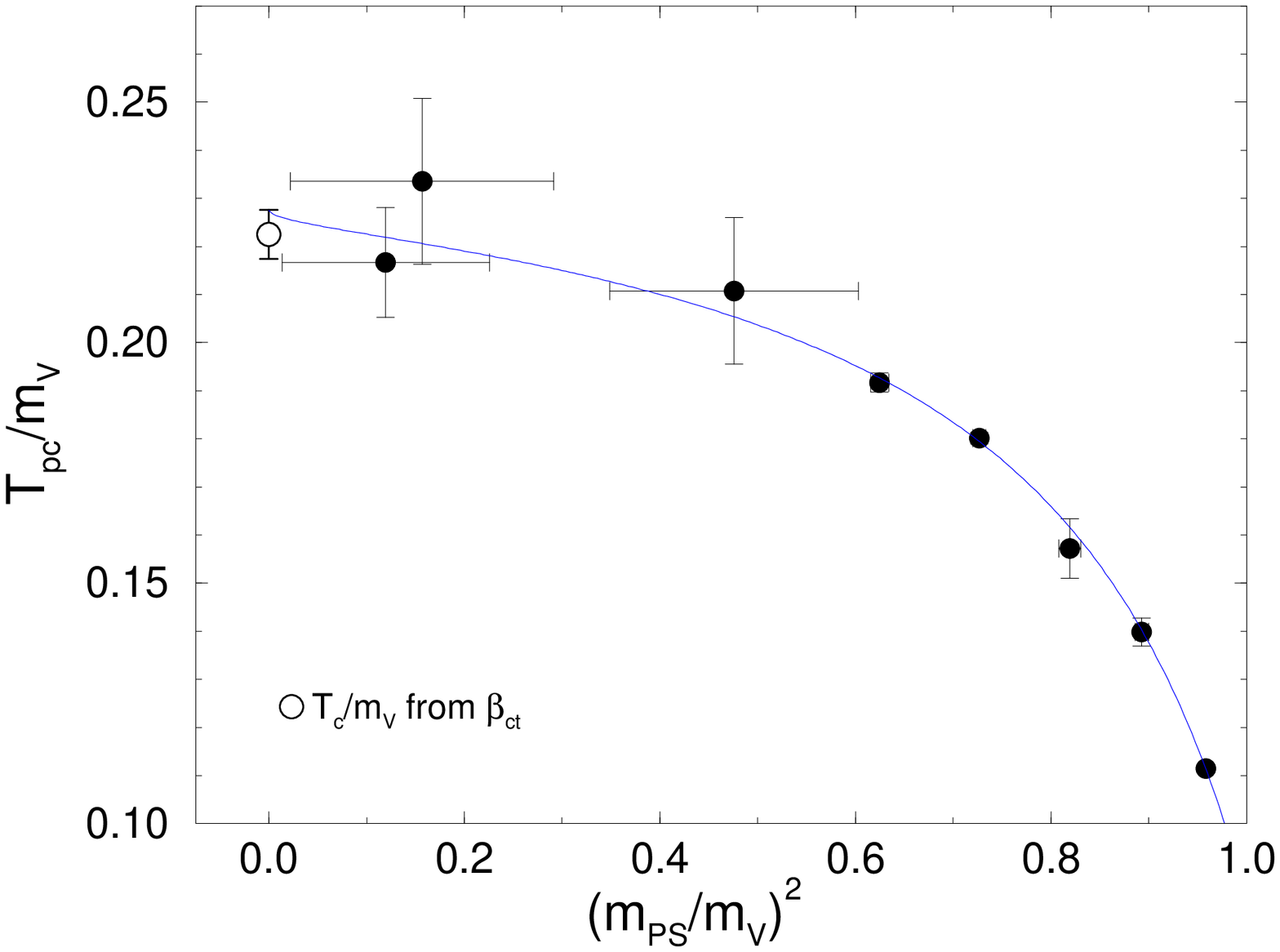} \hspace*{1cm}
\includegraphics[width=7.5cm]{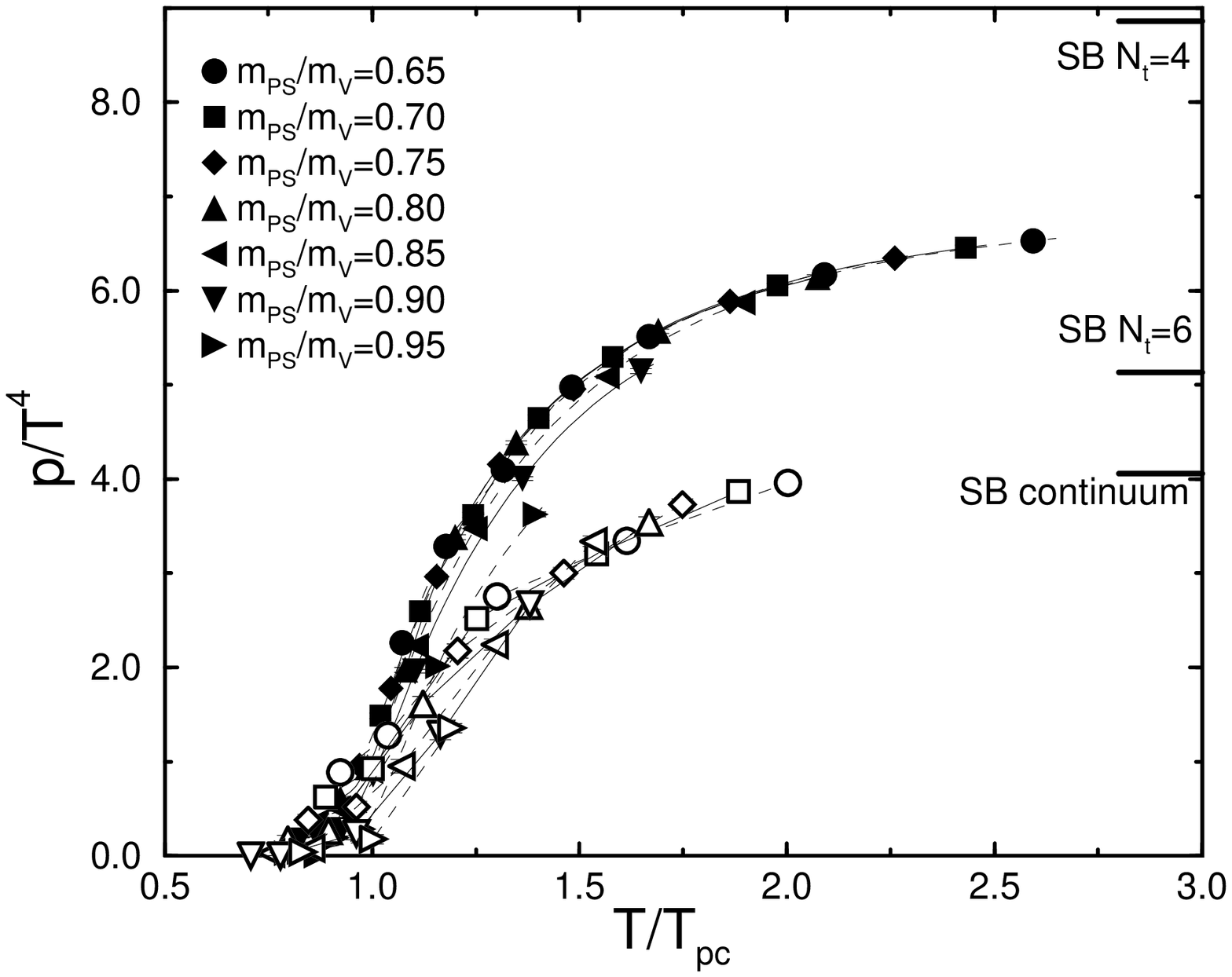}
} \vspace*{-0.6cm}
\caption{Left: the chiral extrapolation of $T_c/m_V$ using improved 
Wilson fermions. Right: the pressure for different quark masses on $N_t=4$
and $N_t=6$ lattices.}
\label{wilson}
\end{figure*}
\subsection{Results with Wilson fermions}
There are numerous results for $T_c$ using Wilson 
quarks~\cite{Bernard:1992yh,Edwards:1999mm,AliKhan:2000iz,Mori:2003pm}.
Fig.~\ref{wilson} (left) shows the chiral extrapolation of $T_c$ 
normalized by the vector boson mass as obtained by the CP-PACS
collaboration~\cite{AliKhan:2000iz} using RG improved gauge- and 
clover improved
Wilson action on $N_t=4$
lattices. The chirally extrapolated result is $T_c=171(4)$~MeV, similarly
to previously discussed staggered results. 

The equation of state as obtained in~\cite{AliKhan:2001ek} is shown in
Fig.~\ref{wilson} (right) for $N_t=4$ and $6$ lattices. 
This analysis was carried out along the LCP.

\subsection{First results with overlap fermions}
As the chiral properties of the action may be important
when we investigate the chiral phase transition in the massless
limit there would be a need to have thermodynamics with
chiral fermions, i.e. with ones satisfying the Ginsparg-Wilson
relation~\cite{Ginsparg:1981bj}.
One possible realization is the Overlap formalism~\cite{Neuberger:1997fp}.
Unfortunately due to the non-analytic form of the Overlap operator,
all simulations are computationally extremely expensive.
In fact, no dynamical simulations have been performed before.

As an exploratory study we implemented the standard hybrid
Monte-Carlo algorithm for the overlap operator~\cite{over_hmc}. The
sign function was approximated by the Zolotarev optimal
rational polynomial approximation~\cite{vandenEshof:2002ms}.
In this approximation one has to perform simultaneous inversions
which is done with a multi-shift solver~\cite{Frommer:1995ik}.
As it was expected, the computational costs are more than two
orders of magnitude higher than for dynamical Wilson fermions.

\begin{figure}
\vspace*{-0.0cm}
\centerline{
\includegraphics[width=7.5cm]{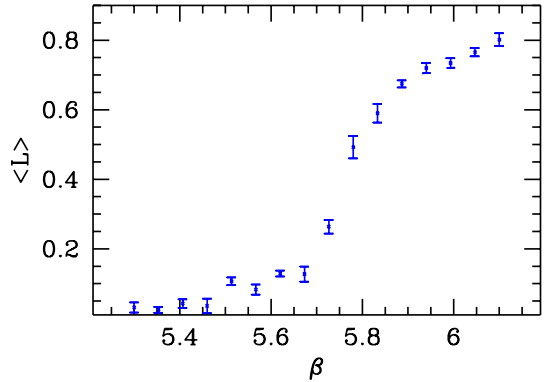}
} \vspace*{-0.8cm}
\caption{The Polyakov loop as a function of the $\beta$ coupling
using dynamical overlap fermions on a $6^3\times 4$ lattice.}
\vspace*{-0.8cm}
\label{overlap}
\end{figure}

The first results on a $6^3\times 4$ lattice are shown in Fig.~\ref{overlap}.
The Polyakov loop shows a clear transition behavior.
This result can of course be considered only as a demonstration
of the working algorithm, for real physical results much larger lattices
with smaller lattice spacings are required.

\subsection{The order of the transition}
As it has been emphasized in the introduction, the order of the
phase transition can only be determined by a careful finite-size
scaling analysis. 
In the case of dynamical simulations these analyses have
given little results yet.
For $N_f=2$ in the massless limit one expects a second order phase
transition in the O(4) universality class. However, only
simulations with improved Wilson quarks could support this 
expectation~\cite{AliKhan:2000iz}. Staggered results do
not seem completely consistent with O(4) 
universality~\cite{Aoki:1998wg,Bernard:1999fv,Engels:2001bq}.

In the case of three degenerate flavors a first order transition
is expected
at zero quark masses. Thus if we increase the quark masses, we 
will have a critical $m_c$ (and correspondingly a critical pseudoscalar mass,
$m_{PSc}$) where the first order transition ends.
This second order endpoint has been located in standard staggered QCD
on $N_t=4$ lattices~\cite{Karsch:2001nf} and the critical mass is 
$m_{PSc} \approx 290$~MeV. However, using improved $p4$ fermions, this
critical mass goes down $m_{PSc} \approx 67$~MeV~\cite{Karsch:2003va} 
indicating that there are still large cutoff effects.
The picture in the physically interesting $N_f=2+1$ case is even more
unclear, however it is very likely that there is only a rapid 
cross-over~\cite{Schmidt:2002uk}.

\subsection{Heavy quark free energy and hadron properties}
One can study the heavy quark free energy at finite temperatures
by measuring the correlation of Polyakov loops:
\be
<L_xL^{\dagger}_y>\sim e^{-V(|x-y|,T)/T}
\ee
As an effect of dynamical quarks, string breaking is expected
at large separations. This string breaking happens at
2-3 times smaller distances as we are getting closer to the critical
temperature~\cite{Bornyakov:2003ev,Kaczmarek:2003dp}. The asymptotic value of the free energy
at large separations shows a strong temperature and quark mass dependence.
As it is expected, for larger temperatures and smaller quark masses the
asymptotic value gets smaller.

Measuring hadron masses is not easy at finite $T$, since the temporal 
extension of the lattice is small. One can either use anisotropic lattices
or measure spatial correlators leading to screening 
masses~\cite{Laermann:2001vg,Gavai:2002jt}.

A recent development is the possibility of reconstructing spectral functions
from the Euclidean correlation functions using the Maximal Entropy 
Method (MEM)~\cite{Asakawa:2000tr}. 
For a good determination, however a large $N_t$
is required.
Quenched results are already available both for
anisotropic lattices up to $N_t=54$ ~\cite{Asakawa:2002xj} and isotropic
lattices up to $N_t=32$~\cite{Karsch:2003wy}.

\section{QCD AT FINITE, BUT SMALL $\mu$}\label{finmu1}
Unlike at finite $T$ but zero $\mu$, at finite chemical potentials
the strategy of lattice simulations is not well defined. The
reason is that at nonzero $\mu$ the fermion determinant becomes
complex which spoils direct importance-sampling based simulations.

Recently several methods have been developed to extract information
for finite $\mu$ from simulations at zero or purely imaginary 
$\mu$ values where the fermion determinant is positive definite.
These techniques are, however still restricted to finite $T$ and relatively
small $\mu$. The validity region is approximately $\mu \lwig T$.
At larger chemical potentials, especially at low temperature, the only
current possibility is to use QCD-like effective models which reflect certain
properties of QCD and which can be solved either exactly or by numerical 
techniques. Some of these models will be discussed in the next section.

\subsection{Multiparameter reweighting}
\begin{figure}
\centerline{
\includegraphics[width=7.5cm]{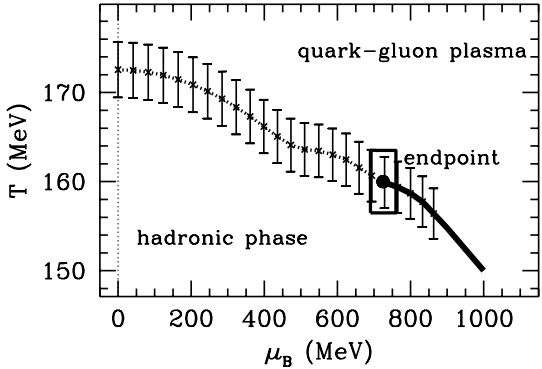}
} \vspace*{-0.6cm}
\caption{The phase diagram of QCD from $N_f=2+1$ simulations
on $N_t=4$ lattices. The dotted part on the left represents
a crossover which is separated from the first-order line
via the critical endpoint.}
\label{endp}
\vspace*{-0.6cm}
\end{figure}
One of the possibilities to extract information at $\mu\neq 0$
is the Glasgow method~\cite{Barbour:1997ej}. It is based on a reweighting
in $\mu$. An ensemble is generated at $\mu=0$ and the ratio of the
fermion determinants at finite $\mu$ and $\mu=0$ is taken into account
as an observable. This method was used to attempt to
locate the phase transition at low $T$ and finite $\mu$, however it fails
even on lattices as small as $4^4$. The reason is the so-called overlap
problem. The generated configurations (only hadronic ones)
do not have enough overlap with the configurations of interest (e.g. 
in the case of a phase transition a mixture of hadronic and quark-gluon
dominated ones).

A simple, but powerful generalization of the Glasgow method is the
overlap improving multiparameter reweighting~\cite{Fodor:2001au}.
The partition function at finite $\mu$ can be rewritten as:
\begin{eqnarray}
\label{rew}
&Z = \int {\cal D}Ue^{-S_{g}(\beta,U)}\det M(m,\mu,U)= \nonumber\\
&\int {\cal D}U e^{-S_{g}(\beta_0,U)}\det M(m_0,\mu=0,U)\\
& \left\{e^{-S_{g}(\beta,U)+S_{g}(\beta_0,U)}
\frac{\det M(m,\mu,U)}{\det M(m_0,\mu=0,U)}\right\} \nonumber,
\end{eqnarray}
where the second line contains a positive definite action
which can be used to generate the configurations and the terms
in the curly bracket in the last line are taken into
account as an observable.
The expectation value of any observable can be then written in the form:
\be
<O>_{\beta,m,\mu}=\frac{\sum O(\beta,m,\mu) w(\beta,m,\mu)}{\sum w(\beta,m,\mu)}
\label{rew_O}
\ee
with $w(\beta,m,\mu)$ being the weights of the configurations defined by
the curly bracket of eqn. (\ref{rew}).

The main difference from the Glasgow method is that reweighting is
done not only in $\mu$ but also in the other parameters of the
action (at least in $\beta$, but possibly also in $m$). This way
the overlap can be improved. If the starting point ($\beta_0, m_0, \mu_0=0$)
is selected to be at the $\mu=0$ transition point then a much better
overlap can be obtained with transition points at higher $\mu$.
One can in general define the best weight lines along which --
starting from a given point ($\beta_0, m_0, \mu_0$) -- the overlap is 
maximal. This can be done e.g. by minimizing the spread of the $w$ weights.
An alternative way is to diagonalize the covariance matrices for infinitesimal
steps~\cite{Crompton:2001ws}.
It has been observed that starting from the $\mu=0$ transition point,
the above defined best weight line coincides with the transition 
line~\cite{Ejiri:2002fj}. 

An interesting question is the goodness of reweighting and its dependence
on the simulation parameters, most importantly on the lattice volumes.
One can define an overlap measure the following way:
let the overlap measure be $2\alpha$ if $\alpha$ fraction of the 
configurations with largest weights (or their absolute values for
complex weights) gives $1-\alpha$ fraction of the total weight.
Clearly for true important sampling, when all the weights are 1, this
definition gives 1 for the overlap measure, otherwise it is smaller.
If we require $2\alpha>1/2$ as a condition for acceptable reweighting,
then the maximal reachable chemical potential, denoted by $\mu_{1/2}$
scales with the volume as 
$\mu_{1/2}\propto V^{-\gamma}$
with $\gamma=0.2-0.3$ from $6^3\times 4, 8^3\times 4, 10^3\times 4$ and 
$12^3 \times 4$ lattice simulations~\cite{eos_long}. 
If $\gamma<0.25$ then it has
the important consequence that the continuum limit can be taken, as
the chemical potential in lattice units we have to reach scales
with $V^{-0.25}$.

\begin{figure}
\centerline{
\includegraphics[width=7.2cm]{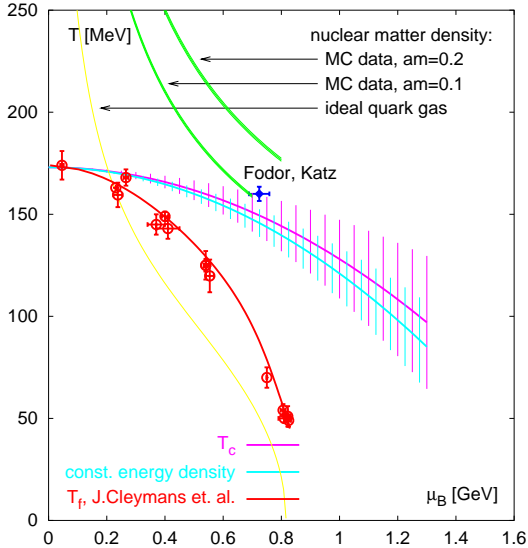}
} \vspace*{-0.6cm}
\caption{The phase diagram of QCD from $N_f=2$ simulations
on $16^3\times 4$ lattices with $p4$ action using Taylor 
expansion~\cite{Allton:2002zi}.}
\label{taylor}
\vspace*{-0.6cm}
\end{figure}

This reweighting technique made it possible to determine the phase diagram 
on the $\mu-T$ plane up to $\mu\approx 1.5T$ in $N_f=4$ ~\cite{Fodor:2001au}
and $N_f=2+1$~\cite{Fodor:2001pe} staggered QCD on $N_t=4$ lattices.
The transition points were located in both cases by finding the
Lee-Yang zeroes of the partition function. 

In the $N_f=2+1$ case, as discussed in the introduction, there is a 
possibility to have a critical endpoint at finite $\mu_E, T_E$ values.
Indeed, the finite size scaling behavior of the imaginary
parts of the Lee-Yang zeroes indicated a cross-over at $\mu=0$
and this turned into a first-order transition at larger $\mu$ values.
After setting the scale with $T=0$ simulations, 
the temperature and baryonic chemical
potential ($\mu_B=3\cdot\mu$) at the endpoint can be given in physical units:
\be
T_E=160 \pm 3.5{\rm MeV},\;\; {\mu_B}_E=725 \pm 35{\rm ~MeV}.
\ee
The phase diagram and the endpoint is shown on Fig.~\ref{endp}.
Note that a similar two-parameter reweighting was succesfully applied
to locate the endpoint of the electroweak phase transition even on large
lattices~\cite{Csikor:1998eu}.

\subsection{Multiparameter reweighting with Taylor-expansion}
The use of eqn. (\ref{rew}) requires the exact calculation
of determinants on each gauge configuration which is computationally
very expensive. 
Instead of using the exact formula, one can make a Taylor expansion
for the determinant ratio in the weights~\cite{Allton:2002zi} (for simplicity
assuming no reweighting in the mass):
\begin{eqnarray}
\ln\left(\frac{\det M(\mu)}{\det M(0)}\right) &=&
\sum_{n=1}^{\infty} \frac{\mu^n}{n!}
\frac{\partial^n \ln \det M(0)}{\partial \mu^n}\nonumber\\
&\equiv&\sum_{n=1}^\infty
{R}_n\mu^n.
\end{eqnarray}       
Taking only the first few terms of the expansion one gets
an approximate reweighting formula. The advantage of
this approximation is that the coefficients are
derivatives of the fermion determinant at $\mu=0$, which 
can be well approximated stochastically.
However, due to the termination of the series and the
errors introduced by the stochastic evaluation of the
coefficients we do not expect this method to work for 
as large $\mu$ values as the full technique.
Indeed, it has been shown in~\cite{deForcrand:2002pa} that
even for very small lattices (i.e. $4^4$) the
phase of the determinant is not reproduced 
by the Taylor expansion for $a\mu>=0.2$.

Fig.~\ref{taylor} shows the phase diagram of $N_f=2$
QCD determined on $16^3\times 4$ lattices with $p4$ action using this 
Taylor-expansion technique~\cite{Allton:2002zi}.
The results are consistent
with the above ones coming from the full technique.
\begin{figure*}
\centerline{
\includegraphics[width=7.0cm]{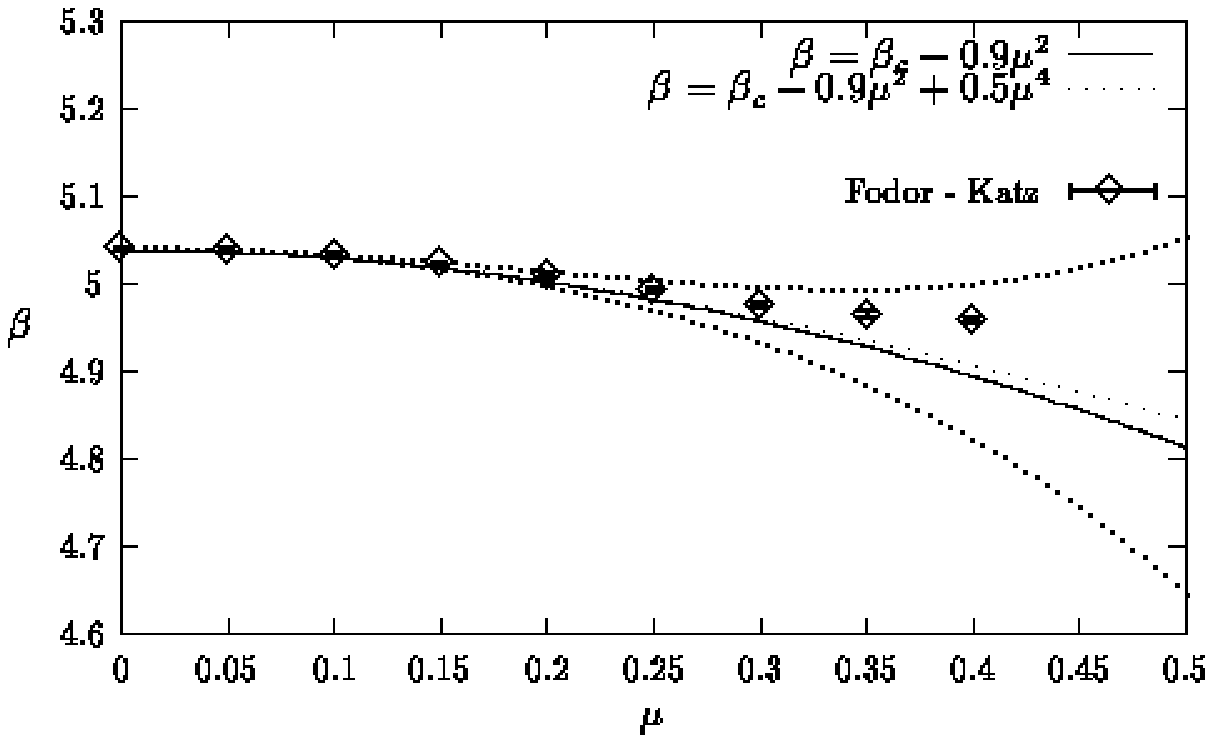} \hspace*{1.7cm}
\includegraphics[width=7.0cm]{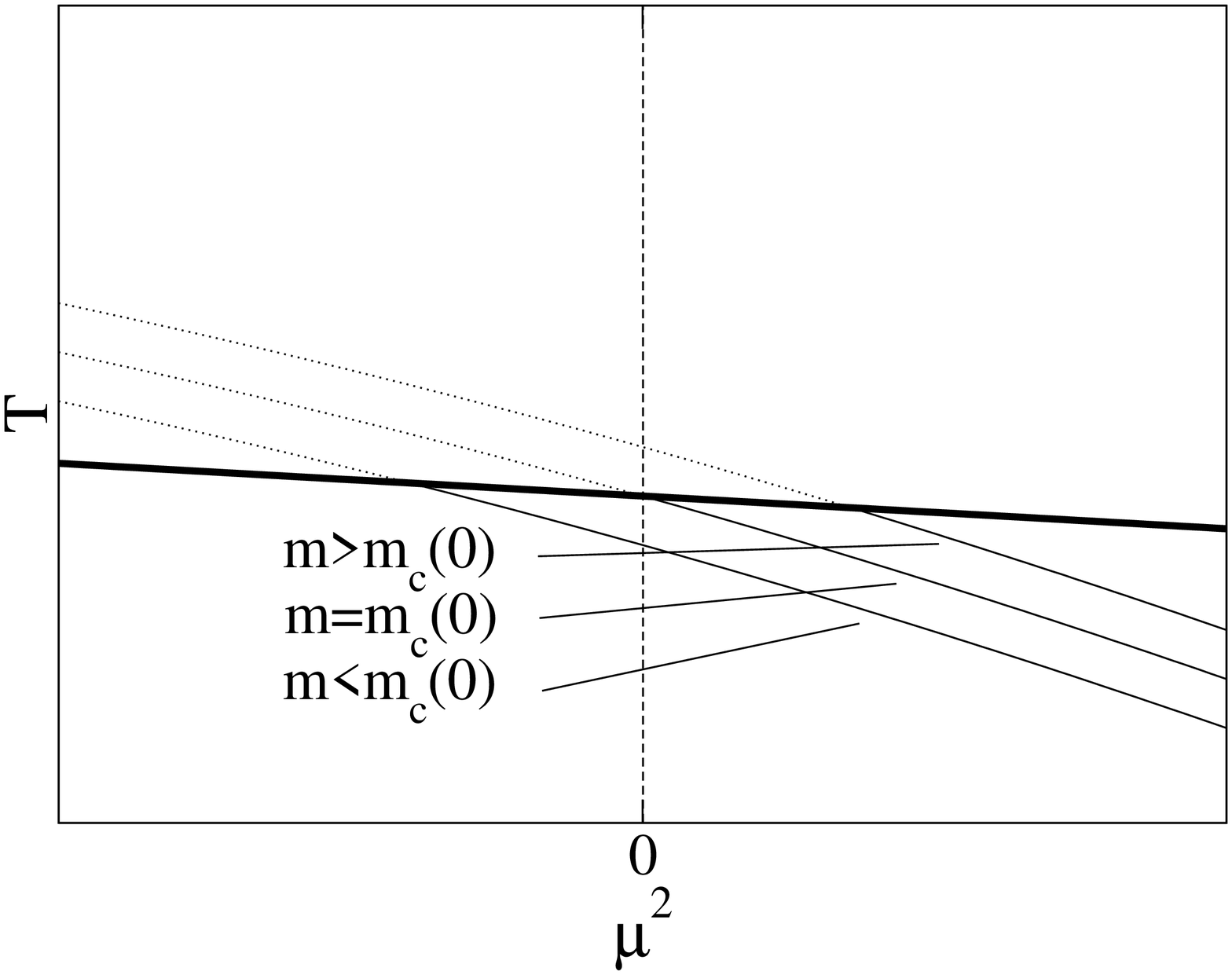}
} \vspace*{-0.6cm}
\caption{Left: the phase diagram of QCD from $N_f=4$ simulations
at imaginary $\mu$ with staggered fermions~\cite{D'Elia:2002gd}.
Right: schematic plot of the phase diagrams (the dotted
parts are cross-over while the solid parts first order regions)
for different quark masses 
and the line of endpoints (thick solid line).}
\label{immu}
\end{figure*}

As discussed in the previous section $N_f=3$ QCD has a critical quark mass
at which there is a second order endpoint at $\mu=0$. If we make reweighting
also in the quark masses then we can keep the system not only on the transition 
line, but also on a critical surface. This way in principle the
critical endpoint at the physical values of the quark masses can be
determined~\cite{Karsch:2003va}. 

\subsection{Simulations at imaginary $\mu$}
The fermion determinant is positive definite if we use a purely imaginary 
chemical potential. So if the transition line $T_c(\mu)$ is an analytic
function then we can determine it for imaginary $\mu$ values and 
analytically continue back to real $\mu$-s. The analytic continuation
is in general impossible from just a finite number of points. However, 
taking a Taylor expansion in $\mu$ or $\mu/T$ one gets:
\be
\frac{T_c(\mu_B)-T_c(0)}{T}=a_2\left(\frac{\mu_B}{T}\right)^2+
a_4\left(\frac{\mu_B}{T}\right)^4+\dots
\ee
The coefficients $a_i$ can be determined from imaginary $\mu$ simulations.
One simply measures $T_c(\mu_I)$ for imaginary $\mu_I$-s and fits it
with a finite order polynomial in $\mu_I/T$.
This method has been applied for $N_f=2$~\cite{deForcrand:2002ci} 
,$N_f=4$~\cite{D'Elia:2002gd} and recently 
for $N_f=3$~\cite{deForcrand:2003hx}
staggered QCD on $N_t=4$ lattices. The $N_f=4$ results are shown on 
Fig.~\ref{immu} (left). The curvatures of the phase diagram in the 
three cases are $a_2$=-0.0056, -0.011 and -0.0068.
The $a_4$ coefficient is very small and
it can only be distinguished from zero with very high 
statistics~\cite{deForcrand:2003hx}.

Similarly to the case of reweighting one can study the mass dependence
of the $\mu_E$ endpoint. In $N_f=3$ QCD for quark masses below the critical
value this endpoint is at imaginary values of $\mu$ (i.e. negative $\mu^2$),
which can be 
investigated. If one assumes the same leading order dependence on $m$
also for real values of $\mu$ (i.e. positive $\mu^2$) one can in principle
determine the location of the critical endpoint for the physical values
of the quark masses (c.f. Fig.~\ref{immu} right).
The first steps have been made in~\cite{deForcrand:2003hx} and the critical
line in the $\mu-m$ plane has been found for 3 degenerate quarks:
\be
\frac{m_c(\mu)}{m_c(\mu=0)}=1+0.84(36)\left(\frac{\mu}{\pi T}\right)^2
\ee

\subsection{The EoS at finite $\mu$}
\begin{figure*}
\centerline{
\includegraphics[bbllx=40,bblly=440,bburx=590,bbury=710,width=15.0cm]{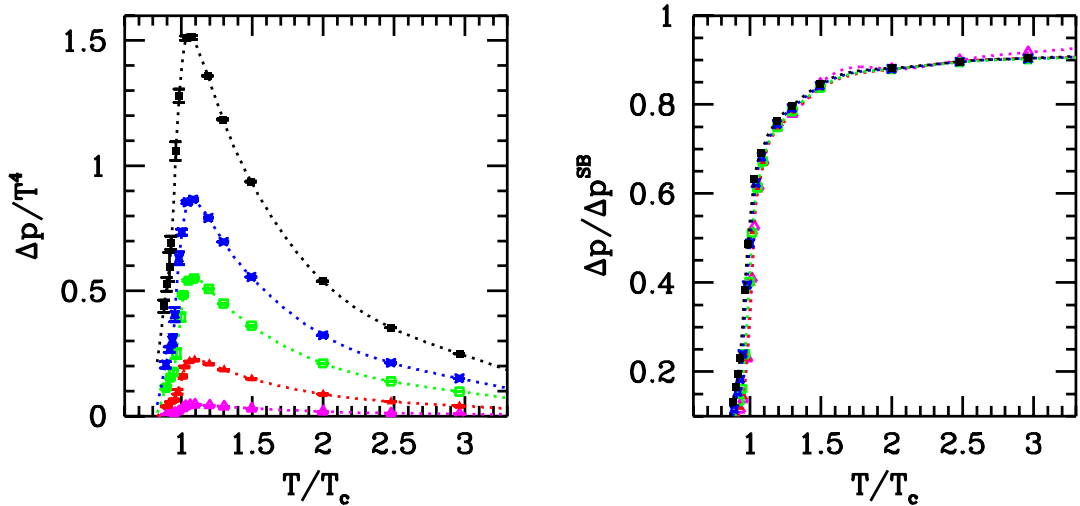}
}  \vspace*{-0.6cm}
\caption{Left: The pressure difference $\Delta p=p(\mu)-p(0)$ for baryonic
chemical potentials $mu_B=100,210,330,410,530$~MeV (from bottom to top) 
obtained with $N_f=2+1$ staggered QCD on $N_t=4$ lattices.
Right: $\Delta p$ normalized by the Stefan-Boltzmann limit.
}
\label{eos}
\end{figure*}
If we perform reweighting not only along the transition line 
but also below or above it, following the above defined best weight lines
then the equation of state can be given at finite $\mu$.
For calculating the pressure or the energy density one simply
has to follow the formulae of the introduction, but all observables
have to be reweighted to finite $\mu$ according to eqn.(\ref{rew_O}).
Calculating the pressure is done by a line integral in the $\beta-\mu$
plane. The simplest choice for the integration path is to start from
deep in the hadronic phase (where we get zero contribution after 
subtracting the $T=0$ contributions) then integrate up to some $\beta_1$
at $\mu=0$ and then follow a best weight line.
                                                                 
Keeping the system always on the LCP is an important aspect of the EoS.
However, if we make reweighting only in $\beta$ and $\mu$ then following the
best weight lines change $\beta$ but not $am_i$, so we automatically leave
the LCP. There are two possible solutions. One could either use also a 
reweighting in $m_i$, which makes all computations more expensive, or
start the reweighting from two different LCP-s and then keep the 
system on the LCP with an appropriate interpolation. The second technique
was used in~\cite{Fodor:2002km} where the EoS was determined for a large
range in $T$ and $\mu$. The results for the pressure are shown in 
Fig.~\ref{eos}. The left panel shows the pressure after subtracting
the $\mu=0$ contribution. On the right panel the pressure is normalized with
the Stefan-Boltzmann limit. We can observe an almost universal scaling 
behavior. After this normalization the pressure
becomes practically independent of $\mu$ in the analyzed region.
One can get similar results using a quasi-particle picture, after
setting the free parameters of the model using $\mu=0$
lattice results~\cite{Szabo:2003kg}.

\begin{figure*}
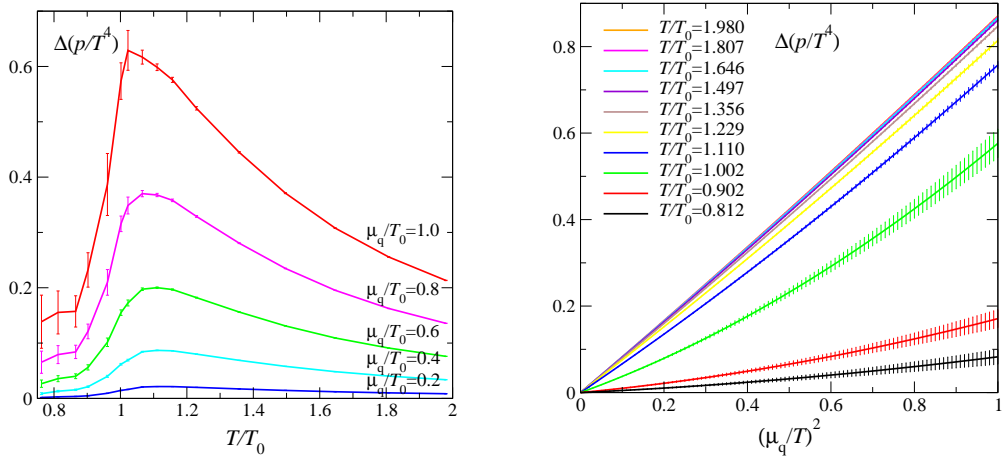

\centerline{\includegraphics*[width=6cm]{eos_taylor2.eps} \hspace*{1cm}
\includegraphics*[width=6cm]{eos_taylor3.eps}
}\vspace*{-0.6cm}
\caption{Left: The pressure difference $\Delta p=p(\mu)-p(0)$ using the
Taylor expansion method with $N_f=2$ staggered QCD with $p4$ action on
$N_t=$ lattices.
Right: $\Delta p$ as a function of $\mu$ for different $T$-s.
}
\label{eos_taylor}
\end{figure*}

The EoS can also be determined using Taylor expansion. The pressure
can be written as:
\be
\frac{p}{T^4}(\mu,T)=
\frac{p_0(T)}{T^4}+c_2\left(\frac{\mu}{T}\right)^2+
c_4\left(\frac{\mu}{T}\right)^4+ \dots
\label{p_taylor}
\ee
The coefficients can be determined with simulations at $\mu=0$.
Fig.~\ref{eos_taylor} shows the results from this 
method~\cite{Allton:2003vx},
which are very similar to those coming from the full technique.
In the vicinity of the critical endpoint the nonlinear susceptibility,
$c_4$ diverges. Thus the convergence radius of the $p(\mu,T)$ function
in principle gives the location of the endpoint. One can estimate
this convergence radius by the ratios of the subsequent 
coefficients.

The pressure has also been computed with the help of 
eqn.(\ref{p_taylor}) using quenched configurations~\cite{Gavai:2003mf}.

Simulations at imaginary $\mu$ can also be used to determine
the $c_i$ coefficients. A first analysis was presented 
in~\cite{D'Elia:2003uy}.

\subsection{The factorization method}
Although it has not yet been tested for QCD, the factorization 
method~\cite{Ambjorn:2002pz} seems very promising, so we discuss it in this 
subsection.
The factorization method was used to study an exactly
solvable random matrix model defined by:
\be
Z=\int dW e^{-N{\rm Tr}(W^\dagger W)}\det D
\ee
with $W$ being $N\times N$ complex matrices and 
\be
D=\left(\begin{array}{cc}m & iW+\mu\\
iW^\dagger+\mu&m\end{array}\right)
\ee
For $\mu\neq 0$ the determinant is complex as in QCD.
The main idea of the factorization method is to
use constrained partition functions:
\be
Z_\nu(x)=\int dW e^{-S}|det D|\delta(x-\nu)
\ee
with some real-valued 
observable $\nu$. If one performs
simulations at several $x$ values using $Z_\nu(x)$ and includes
the phase of the determinant as an observable then after integrating
over $x$ one can get $<\nu>$ for the full theory. The advantage of this 
approach is that each value of the observable is sampled with the help 
of $Z_\nu(x)$ thus solving the overlap problem. One disadvantage is
of course the need of simulations at several $x$ values and that for each 
observable an independent simulation is needed. The implementation
of $Z_\nu(x)$ for QCD especially for the most interesting fermionic 
observables (e.g. quark density) is also not trivial.
However, as this method produced results completely consistent with the 
exact ones for this simple random matrix theory~\cite{Ambjorn:2002pz},
we could expect similar success for QCD.
The technique is not expected to work at large $\mu$ because 
a reweighting is still needed to include the phases of the determinant
and when they become strongly oscillating this may make the measurement
of the observables even with the constrained partition functions impossible.

\section{QCD-LIKE MODELS}\label{finmu2}
At low temperature and large densities a rich phase structure of QCD
is conjectured. Due to asymptotic freedom at large
densities quarks are expected to be weakly coupled. However, as there
is still a weak attractive interaction they can form Cooper-pairs
and thus break the SU(3) symmetry spontaneously and lead to 
color superconductivity. It is also expected that the ground state
will be a color-flavor locked (CFL) phase.

Unfortunately these regions are
still unavailable for lattice simulations due to the overlap problem
and the strong oscillation of the phase of the fermion determinant.
One can, however study QCD-like models, which have positive actions and
which are expected to reproduce some features of high density QCD.
In the following we briefly discuss the most widely used models to study
high density phenomena.
\begin{figure}
\centerline{
\includegraphics*[width=7cm]{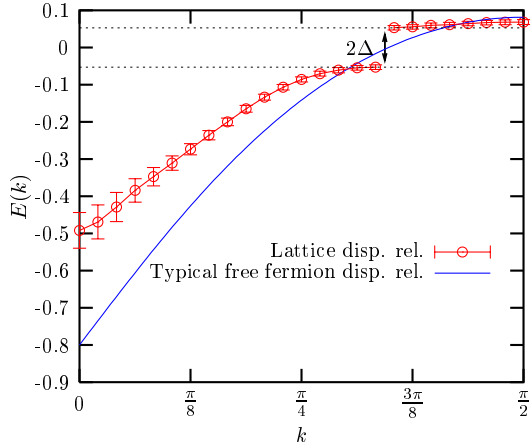}}
\caption{The energy-momentum dispersion relation for the 3+1
dimensional NJL model. The energy
gap can be seen clearly. 
} \vspace*{-0.6cm}
\label{NJL}
\end{figure}

\subsection{Two color QCD}
QCD with two colors has a real fermion determinant, which is also
positive definite for an even number of flavors even at nonzero chemical 
potential. Thus its properties can be studied on the whole $\mu-T$ plane
with standard Monte-Carlo methods.
Phase transitions have been observed both at finite $T$ and $\mu=0$ and
at low $T$ and finite $\mu$~\cite{Hands:2000ei,Kogut:2002cm}. 
An interesting property of two-color QCD that it also has an
endpoint in the $\mu-T$ plane so it can be used to test any method
which is used to locate the endpoint of real QCD.

Unfortunately there is no Fermi surface in this model, so it can not be used
to study superconductivity.

The modification of hadron properties due to the chemical potential
has also been investigated in this model~\cite{Muroya:2002ry}.

\subsection{Isospin chemical potential}
If we introduce opposite chemical potentials for the $u$ and $d$ 
quarks then the phases of their fermion determinants cancel, so 
QCD with such an isospin $\mu$ has a positive definite action.
This model has been studied in detail recently~\cite{Kogut:2002zg}.
Interestingly at low densities it behaves very similarly to QCD
with real chemical potential, which leads to the conclusion that
at finite $T$ and low density the phase of the fermion determinant
does not play an important role.
On the other hand, at low temperatures, just as expected, a
phase transition happens at $\mu\approx m_\pi/2$, as the pions 
are the lowest excitations in this system. This transition should
not be present in QCD with baryonic $\mu$. 

As in the case of two-color QCD, Fermi phenomena can not be studied in this 
model.

\subsection{Four-fermion models}
In principle one could integrate out the gauge fields in QCD leading
to en effective purely fermionic theory.
The first non-trivial interaction term of such an effective theory
is a four-fermion interaction. 
Thus, models with four-fermion interactions are expected to reflect some
features of QCD.

The 2+1 dimensional Gross-Neveu model has been studied in~\cite{Hands:2003dh}.
Fermi phenomena have been observed, however no gap was found in the 
energy-momentum dispersion relations.
The non-exponential decay of meson correlation functions indicate
the existence of massless particle-hole excitations.

The NJL model was studied both in 2+1~\cite{Hands:2000gv} and 
3+1~\cite{Hands:2002mr} 
dimensions. Interestingly due to the different dimensionalities the two 
models behave differently in the large density regions. While in both
models a $qq$ pairing has been observed, only the latter system shows
BCS condensation and an energy gap in the dispersion relation 
(see Fig.~\ref{NJL}). 
Unfortunately this 3+1 dimensional model is not renormalizable, 
however using a finite cutoff, one can set the parameters of the theory 
to describe low-energy observables of QCD.

\subsection{Low energy effective models}
Although the action of QCD at large densities is complex, it has been shown
in~\cite{Hong:2003zq} that if one is only interested in the low energy 
behavior, i.e. modes close to the Fermi surface, then a low-energy
effective action can be derived which is positive definite (see 
also~\cite{Hands:2003rw}). Using 
this low energy effective action it has been proven that the CFL phase
is the ground state of QCD at large densities.

Another approach is to build up an effective Hamiltonian based on 
strong coupling expansion, which can then be used to study phenomena even
in the chiral limit~\cite{Fang:2002rk}.

\section{CONCLUSIONS}\label{conc}
In this review recent results of lattice QCD at finite $T$ and/or $\mu$
have been discussed.
For the pure gauge theory there are already continuum-extrapolated
predictions both for the phase transition temperature and the equation
of state. The transition is proven to be first order.

Including dynamical quarks makes all computations much more complicated.
Therefore there are still no   chiral- and continuum-extrapolated results.
However, the transition temperature, the equation of state, the heavy quark
potential and hadron properties are quite well known and all necessary,
but still missing extrapolations can be expected in the near future.

The case of finite $\mu$ is even more complicated due to the complexness
of the action. The exploration 
of the phase diagram, the critical endpoint and
the equation of state has just began recently. Lots of new results can be
expected in the near future especially in the small $\mu$ region.

Unfortunately even the latest methods are still incapable of solving the 
complex action problem which is most severe for the very interesting large 
$\mu$, small $T$ region. Therefore effective, QCD-like models are used
to investigate this parameter region. These effective models seem to
support the assumption on the existence of a color-superconducting
phase at large densities.

\section{ACKNOWLEDGMENTS}
I thank Zolt\'an Fodor for many useful discussions and 
for careful reading of the manuscript.


\begin{thebibliography}{9}

\bibitem{McLerran:1980pk}
L.~D.~McLerran and B.~Svetitsky,
Phys.\ Lett.\ B {\bf 98} (1981) 195.

\bibitem{Kuti:1980gh}
J.~Kuti, J.~Polonyi and K.~Szlachanyi,
Phys.\ Lett.\ B {\bf 98} (1981) 199.

\bibitem{Yang:be}
C.~N.~Yang and T.~D.~Lee,
Phys.\ Rev.\  {\bf 87} (1952) 404;
Phys.\ Rev.\  {\bf 87} (1952) 410.

\bibitem{Stephanov:1998dy}
M.~A.~Stephanov, K.~Rajagopal and E.~V.~Shuryak,
Phys.\ Rev.\ Lett.\  {\bf 81} (1998) 4816.

\bibitem{Boyd:1996bx}
G.~Boyd, J.~Engels, F.~Karsch, E.~Laermann, C.~Legeland, M.~Lutgemeier and B.~Petersson,
Nucl.\ Phys.\ B {\bf 469} (1996) 419.

\bibitem{Beinlich:1997ia}
B.~Beinlich, F.~Karsch, E.~Laermann and A.~Peikert,
Eur.\ Phys.\ J.\ C {\bf 6} (1999) 133.

\bibitem{Okamoto:1999hi}
M.~Okamoto {\it et al.}  [CP-PACS Collaboration],
Phys.\ Rev.\ D {\bf 60} (1999) 094510.

\bibitem{Namekawa:2001ih}
Y.~Namekawa {\it et al.}  [CP-PACS Collaboration],
Phys.\ Rev.\ D {\bf 64} (2001) 074507.


\bibitem{Sommer:1993ce}
R.~Sommer,
Nucl.\ Phys.\ B {\bf 411} (1994) 839.

\bibitem{Necco:2003vh}
S.~Necco,
hep-lat/0309017.

\bibitem{Fukugita:1989yb}
M.~Fukugita, M.~Okawa and A.~Ukawa,
Phys.\ Rev.\ Lett.\  {\bf 63} (1989) 1768.

\bibitem{Karsch:2000kv}
F.~Karsch, E.~Laermann and A.~Peikert,
Nucl.\ Phys.\ B {\bf 605} (2001) 579.

\bibitem{Bernard:2003dk}
C.~Bernard {\it et al.}  [MILC Collaboration],
hep-lat/0309118.


\bibitem{Gavai:2002fi}
R.~V.~Gavai and S.~Gupta,
Phys.\ Rev.\ D {\bf 66} (2002) 094510.

\bibitem{Bernard:1996cs}
C.~W.~Bernard {\it et al.}  [MILC Collaboration],
Phys.\ Rev.\ D {\bf 55} (1997) 6861
hep-lat/9612025].

\bibitem{Karsch:2000ps}
F.~Karsch, E.~Laermann and A.~Peikert,
Phys.\ Lett.\ B {\bf 478} (2000) 447.

\bibitem{eos_long}
F.~Csikor, G.I.~Egri, Z.~Fodor, S.D.~Katz, K.K.~Szabo, A.I.~Toth,
in preparation.

\bibitem{Bernard:1992yh}
C.~W.~Bernard {\it et al.},
Phys.\ Rev.\ D {\bf 46} (1992) 4741.

\bibitem{Edwards:1999mm}
R.~G.~Edwards and U.~M.~Heller,
Phys.\ Lett.\ B {\bf 462} (1999) 132.

\bibitem{AliKhan:2000iz}
A.~Ali Khan {\it et al.}  [CP-PACS Collaboration],
Phys.\ Rev.\ D {\bf 63} (2001) 034502.

\bibitem{Mori:2003pm}
Y.~Mori {\it et al.},
Nucl.\ Phys.\ A {\bf 721} (2003) 930
hep-lat/0301003].

\bibitem{AliKhan:2001ek}
A.~Ali Khan {\it et al.}  [CP-PACS collaboration],
Phys.\ Rev.\ D {\bf 64} (2001) 074510.

\bibitem{Ginsparg:1981bj}
P.~H.~Ginsparg and K.~G.~Wilson,
Phys.\ Rev.\ D {\bf 25} (1982) 2649.

\bibitem{Neuberger:1997fp}
H.~Neuberger,
Phys.\ Lett.\ B {\bf 417} (1998) 141.

\bibitem{over_hmc}
Z.~Fodor, S.D.~Katz and K.K. Szabo,

in preparation.

\bibitem{vandenEshof:2002ms}
J.~van den Eshof, A.~Frommer, T.~Lippert, K.~Schilling and H.~A.~van der Vorst,
Comput.\ Phys.\ Commun.\  {\bf 146} (2002) 203;
T.~W.~Chiu, T.~H.~Hsieh, C.~H.~Huang and T.~R.~Huang,
Phys.\ Rev.\ D {\bf 66} (2002) 114502.

\bibitem{Frommer:1995ik}
A.~Frommer, B.~Nockel, S.~Gusken, T.~Lippert and K.~Schilling,
Int.\ J.\ Mod.\ Phys.\ C {\bf 6} (1995) 627.

\bibitem{Aoki:1998wg}
S.~Aoki {\it et al.}  [JLQCD Collaboration],
Phys.\ Rev.\ D {\bf 57} (1998) 3910.

\bibitem{Bernard:1999fv}
C.~W.~Bernard {\it et al.},
Phys.\ Rev.\ D {\bf 61} (2000) 054503.

\bibitem{Engels:2001bq}
J.~Engels, S.~Holtmann, T.~Mendes and T.~Schulze,
Phys.\ Lett.\ B {\bf 514} (2001) 299.


\bibitem{Karsch:2001nf}
F.~Karsch, E.~Laermann and C.~Schmidt,
Phys.\ Lett.\ B {\bf 520} (2001) 41.

\bibitem{Karsch:2003va}
F.~Karsch, C.~R.~Allton, S.~Ejiri, S.~J.~Hands, O.~Kaczmarek, E.~Laermann and C.~Schmidt,
hep-lat/0309116.

\bibitem{Schmidt:2002uk}
C.~Schmidt, C.~R.~Allton, S.~Ejiri, S.~J.~Hands, O.~Kaczmarek, F.~Karsch and E.~Laermann,
hep-lat/0209009.

\bibitem{Bornyakov:2003ev}
V.~Bornyakov {\it et al.},
hep-lat/0301002.

\bibitem{Kaczmarek:2003dp}
O.~Kaczmarek, F.~Karsch, P.~Petreczky and F.~Zantow,
hep-lat/0309121.

\bibitem{Laermann:2001vg}
E.~Laermann and P.~Schmidt,
Eur.\ Phys.\ J.\ C {\bf 20} (2001) 541.

\bibitem{Gavai:2002jt}
R.~V.~Gavai and S.~Gupta,
Phys.\ Rev.\ D {\bf 67} (2003) 034501.

\bibitem{Asakawa:2000tr}
M.~Asakawa, T.~Hatsuda and Y.~Nakahara,
Prog.\ Part.\ Nucl.\ Phys.\  {\bf 46} (2001) 459.

\bibitem{Asakawa:2002xj}
M.~Asakawa, T.~Hatsuda and Y.~Nakahara,
Nucl.\ Phys.\ A {\bf 715} (2003) 863.

\bibitem{Karsch:2003wy}
F.~Karsch, E.~Laermann, P.~Petreczky and S.~Stickan,
Phys.\ Rev.\ D {\bf 68} (2003) 014504;
P.~Petreczky, S.~Datta, F.~Karsch and I.~Wetzorke,
hep-lat/0309012.

\bibitem{Barbour:1997ej}
I.~M.~Barbour, S.~E.~Morrison, E.~G.~Klepfish, J.~B.~Kogut and M.~P.~Lombardo,
Nucl.\ Phys.\ Proc.\ Suppl.\  {\bf 60A} (1998) 220.

\bibitem{Fodor:2001au}
Z.~Fodor and S.~D.~Katz,
Phys.\ Lett.\ B {\bf 534} (2002) 87.

\bibitem{Crompton:2001ws}
P.~R.~Crompton,
Nucl.\ Phys.\ B {\bf 619} (2001) 499.

\bibitem{Ejiri:2002fj}
S.~Ejiri,
hep-lat/0212022.



\bibitem{Fodor:2001pe}
Z.~Fodor and S.~D.~Katz,
JHEP {\bf 0203} (2002) 014.

\bibitem{Csikor:1998eu}
F.~Csikor, Z.~Fodor and J.~Heitger,
Phys.\ Rev.\ Lett.\  {\bf 82} (1999) 21.

\bibitem{Allton:2002zi}
C.~R.~Allton {\it et al.},
Phys.\ Rev.\ D {\bf 66} (2002) 074507.

\bibitem{deForcrand:2002pa}
P.~de Forcrand, S.~Kim and T.~Takaishi,
hep-lat/0209126.

\bibitem{deForcrand:2002ci}
P.~de Forcrand and O.~Philipsen,
Nucl.\ Phys.\ B {\bf 642} (2002) 290.

\bibitem{D'Elia:2002gd}
M.~D'Elia and M.~P.~Lombardo,
Phys.\ Rev.\ D {\bf 67} (2003) 014505.

\bibitem{deForcrand:2003hx}
P.~de Forcrand and O.~Philipsen,
hep-lat/0307020.


\bibitem{Fodor:2002km}
Z.~Fodor, S.~D.~Katz and K.~K.~Szabo,
Phys.\ Lett.\ B {\bf 568} (2003) 73.

\bibitem{Szabo:2003kg}
K.~K.~Szabo and A.~I.~Toth,
JHEP {\bf 0306} (2003) 008.


\bibitem{Allton:2003vx}
C.~R.~Allton, S.~Ejiri, S.~J.~Hands, O.~Kaczmarek, F.~Karsch, E.~Laermann and C.~Schmidt,
Phys.\ Rev.\ D {\bf 68} (2003) 014507.

\bibitem{Gavai:2003mf}
R.~V.~Gavai and S.~Gupta,
Phys.\ Rev.\ D {\bf 68} (2003) 034506.

\bibitem{D'Elia:2003uy}
M.~D'Elia and M.~P.~Lombardo,
hep-lat/0309114.



\bibitem{Ambjorn:2002pz}
J.~Ambjorn, K.~N.~Anagnostopoulos, J.~Nishimura and J.~J.~Verbaarschot,
JHEP {\bf 0210} (2002) 062.

\bibitem{Hands:2000ei}
S.~Hands, I.~Montvay, S.~Morrison, M.~Oevers, L.~Scorzato and J.~Skullerud,
Eur.\ Phys.\ J.\ C {\bf 17} (2000) 285.

\bibitem{Kogut:2002cm}
J.~B.~Kogut, D.~Toublan and D.~K.~Sinclair,
Nucl.\ Phys.\ B {\bf 642} (2002) 181.

\bibitem{Muroya:2002ry}
S.~Muroya, A.~Nakamura and C.~Nonaka,
Phys.\ Lett.\ B {\bf 551} (2003) 305.

\bibitem{Kogut:2002zg}
J.~B.~Kogut and D.~K.~Sinclair,
Phys.\ Rev.\ D {\bf 66} (2002) 034505.

\bibitem{Hands:2003dh}
S.~Hands, J.~B.~Kogut, C.~G.~Strouthos and T.~N.~Tran,
Phys.\ Rev.\ D {\bf 68} (2003) 016005.

\bibitem{Hands:2000gv}
S.~Hands, B.~Lucini and S.~Morrison,
Phys.\ Rev.\ Lett.\  {\bf 86} (2001) 753.

\bibitem{Hands:2002mr}
S.~Hands and D.~N.~Walters,
Phys.\ Lett.\ B {\bf 548} (2002) 196; hep-lat/0308030.

\bibitem{Hong:2003zq}
D.~K.~Hong and S.~D.~Hsu,
Phys.\ Rev.\ D {\bf 68} (2003) 034011.

\bibitem{Hands:2003rw}
S.~Hands,
hep-ph/0310080.

\bibitem{Fang:2002rk}
Y.~Z.~Fang and X.~Q.~Luo,
hep-lat/0210031.





\end{thebibliography}
\end{document}